\documentclass[twocolumn,showpacs,preprintnumbers,nofootinbib,amsmath,amssymb]{revtex4}

\begin{document}

\title{Superstring in doubled superspace}

\author{Igor Bandos
$^{\dagger\diamondsuit}$
}
\address{
$^{\dagger}$
Department of
Theoretical Physics, University of the Basque Country UPV/EHU,
P.O. Box 644, 48080 Bilbao, Spain
 \\
 $^{\diamondsuit}$
IKERBASQUE, Basque Foundation for Science, 48011, Bilbao, Spain
}

\date{July 28, 2015. V2: October 7, 2015, Printed \today}

\begin{abstract}

The covariant and kappa--symmetric action for superstring in direct product of two flat $D=10$ ${\cal N}=1$ superspaces is presented. It is given by the sum of supersymmetric generalization of two copies of chiral boson actions constructed with the use of the Pasti-Sorokin-Tonin (PST) technique. The chirality of 8 `left' bosons and 8 `left' fermions and the anti-chirality of their `right' counterparts are obtained as gauge fixed version of the equations of motion, so that the physical degrees of freedom are essentially those of the  II Green-Schwarz superstring. Our action is manifestly T-duality invariant as the fields describing oscillating and winding modes enter it on equal footing.

\end{abstract}


\maketitle

\section{Introduction}

The fact that, to make manifest  T-duality, the characteristic symmetry of String Theory \cite{Green:1987sp}, one needs to double the coordinates of the spacetime had been  appreciated decades ago and the corresponding sigma model actions were  considered in \cite{Duff:1989tf} and
\cite{Tseytlin:1990nb}. The background fields coupled to the fundamental string  should also allow for formulation in doubled spacetime (or should be doubled themselves \cite{Siegel:1993xq}); such a description is now known as 'double field theory' \cite{Hull:2009mi}.

The pseudoaction of sigma model on doubled twisted tori was proposed
in \cite{Hull:2004in} (where 'pseudo' refers to the fact that a chirality/self-duality condition have to be imposed by hand when working with the action) while Tseytlin's action for this case was studied in \cite{Dall'Agata:2008qz}. The covariant version of Tseytlin's action  was constructed in  \cite{Berman:2007xn} with the use of Pasti-Sorokin-Tonin (PST) technique \cite{Pasti:1995tn}  (see also \cite{Copland:2011wx}). The relation of the covariant pseudo-action \cite{Hull:2004in} and
non-covariant action \cite{Tseytlin:1990nb} was the subject of \cite{DeAngelis:2013wba}, where the PST action of  \cite{Berman:2007xn} was also reproduced .  However, neither a supersymmetric generalization of the PST-type covariant action from \cite{Berman:2007xn} nor a supersymmetric generalization of the non-covariant Tseytlin's action \cite{Tseytlin:1990nb} have been known before.

A superstring model  in an enlarged doubled superspace was proposed in recent  \cite{Hatsuda:2015cia} (see also \cite{Hatsuda:2014qqa}). This model is not just the supersymmetric and covariant generalization of Tseytlin's action as a number of additional types of coordinates are introduced besides the doubling of bosonic vector coordinates of superspace. The approach of \cite{Hatsuda:2015cia} is based on an embedding of (two) super-Poincar\'e algebra(s) into a superalgebra with nondegenerate metric; extra coordinates are  then  partially removed by imposing dimensional reduction and section conditions.

In this letter we develop a more economic description of the T-duality invariant type II superstring. We
present the PST action for supersymmetric string in double $D=10$, ${\cal N}=1$ superspace   
and describe its gauge symmetries, including the local fermionic  $\kappa$--symmetry. 
Our action is a supersymmetric generalization of the  bosonic string PST action from 
\cite{DeAngelis:2013wba} and the nontrivial part of our study was the search for $\kappa$-symmetry. 
The preservation of the $\kappa$--symmetry may be considered as a guide to search for superspace 
constraints describing supergravity in doubled superspace (see \cite{Coimbra:2011nw,Jeon:2011vx} 
for the description of supergravity in double space
\footnote{Is M-theoretic counterparts in enlarged spaces  with manifest U-duality symmetry 
 were studied in \cite{Coimbra:2012af} and \cite{Godazgar:2014nqa,Musaev:2014lna,Abzalov:2015ega}; the their purely bosonic counterparts had been proposed in  
\cite{Hull:2007zu} and  \cite{Hohm:2013pua}.} and \cite{Hatsuda:2014aza,Linch:2015fca} for its description in a very extended doubled superspace).

We denote the coordinates of two copies, left, $\Sigma_L^{(10|16)}$,  and right, $\Sigma_R^{(10|16)}$, of the  $D=10$, ${\cal N}=1$ superspace  by
\begin{eqnarray}
\label{cZL=}
{\cal Z}_L^M = (X^a_L,\theta^{\alpha 1})\qquad and \qquad {\cal Z}_R^{\tilde{M}} = (X^{\tilde{a}}_R,\theta^{\tilde{\alpha} 2})\qquad
\end{eqnarray}
with $a=0,1,...,9$ and $\alpha=1,...,16$. Their push-forwards to the worldsheet $W^2$ (with local coordinates  $\{\xi^m\}$),  which we will denote by the same symbols,  will describe the left-moving and right-moving modes of type II superstring. The upper position of the index of $\theta^{\tilde{\alpha} 2}$ does not mean that we are restricting ourselves to IIB superstring as far as the indices of left and right coordinates are transformed by different $SO(1,9)$ groups. This is reflected by tilde over the right coordinate indices (we omit this below to simplify the notation). The  $SO(1,9)\times SO(1,9)$ will be a manifest symmetry of our superstring action.

In very early papers \cite{Isaev:1988qn} Isaev and Ivanov showed that the standard Green-Schwarz (GS) superstring action \cite{Green:1983wt}  can be written as an action on
 $\Sigma_L^{(10|16)}\oplus \Sigma_R^{(10|16)}$.
 However in this case only the sum of two 10D vector coordinates, $X^a= X_L^a+X_R^a$,  is really present while the difference, $\tilde{X}{}^a= X_L^a-X_R^a$, enters the Lagrangian form under total derivative only.
Indeed, the  Volkov-Akulov (VA) 1-form of the type II superspace  $\Sigma_L^{(10|16+16)}$,
$\Pi^a=dX^a  - i d\theta^1\sigma^a\theta^1  - i d\theta^2\sigma^a\theta^2 $, which is used to construct the first, Nambu-Goto type term of the GS action,  can be considered as a sum $\Pi^a=L^a+R^a$, of  VA 1-forms of left and right superspaces,
\begin{eqnarray}
\label{La=}
L^a =dX^a_L - i d\theta^1\sigma^a\theta^1 \; , \qquad R^a =dX^a_R - i d\theta^2\sigma^a\theta^2 \; , \qquad
\end{eqnarray}
while the second Wess--Zumino term of the GS action can be written as $\int (L^a+R^a) \wedge d(L^a-R^a)$. In this form  $\tilde{X}{}^a= X_L^a-X_R^a$, although appearing in  the building blocks, disappear from the action ($dd=0$), while  ${X}{}^a= X_L^a+X_R^a$ obey second order  equations. In a suitable gauge these equations for transverse, physical components ($a=I=1,...,8$) are solved by the sum of chiral (left--moving) and anti-chiral (right--moving) functions,  $\tilde{x}_L^I(\xi^0 +\xi^1) $ and $\tilde{x}_R^I(\xi^0 -\xi^1)$; generically these do not  coincide  with above  $X_L^a$ and $X_R^a$.

In contrast, our action below is formulated in terms of originally unconstrained
${\cal Z}_L^M(\xi)$ and ${\cal Z}_R^{\tilde{M}}(\xi)$ which  become (essentially) chiral and anti-chiral (self-dual and anti-self-dual) on the mass shell after gauge fixing one of the gauge symmetries (PST$_1$ symmetry) of the action.

\section{Action for superstring in doubled superspace}

Our T-duality invariant type II superstring action is given by the sum
\begin{eqnarray}
\label{S=SL+SR}
S= S_L +S_R= \int_{W^2} {\cal L}_{2}^L[{\cal Z}_L (\xi)] + \int_{W^2} {\cal L}_{2}^R[{\cal Z}_R(\xi)]
\end{eqnarray}
of two basically independent actions for `left' and `right' supermultiplets of coordinate fields ${\cal Z}_L^M (\xi) = (X^a_L(\xi),\theta^{\alpha 1}(\xi))$ and
${\cal Z}_R^{\tilde{M}}(\xi) = (X^{\tilde{a}}_R(\xi) ,\theta^{\tilde{\alpha} 2}(\xi))$
corresponding to $\Sigma^{(10|16)}_L$ and  $\Sigma^{(10|16)}_R$ coordinates (\ref{cZL=}); hence the name of superstring in double superspace. We denote
 local coordinates on the worldsheet $W^2$ by $\xi^m=(\xi^0,\xi^1)=(\tau, \sigma)$.
The Lagrangian 2-form of the left supermultiplet, ${\cal L}_{2}^L$ in (\ref{S=SL+SR}),  in its turn is given by the sum
\begin{eqnarray}
\label{SL=}
 {\cal L}_{2}^L ={\cal L}_{2}^{PST}+  {\cal L}_{2}^{WZ}\;
\end{eqnarray}
of the PST Lagrangian form
\begin{eqnarray}
\label{L=PST}
 && {\cal L}_{2}^{PST}= L^a \wedge v \, i_v(L_a -*L_a)\nonumber \\ && \quad \equiv  {1\over 2} *L_a\wedge L^a  + {1\over 2}
  i_v(L_a -*L_a) \, * i_v(L^a -*L^a)\;  \qquad
\end{eqnarray}
and of the Wess--Zumino (WZ) term
\begin{eqnarray}
\label{L=WZ}
{\cal L}_{2}^{WZ} = - i dX_L^a \wedge  d\theta^1\sigma_a\theta^1 \; .
\end{eqnarray}
Here  $\wedge$ is the exterior product which is
antisymmetric for bosonic 1--forms \footnote{so that, e.g.
 $dX_L^a \wedge dX_L^b=- dX_L^b  \wedge dX_L^a$ and $dX_L^a \wedge  d\theta^1=-
 d\theta^1 \wedge dX_L^a$, but $d\theta^{\alpha 1} \wedge d\theta^{\beta 1}= +  d\theta^{\beta 1}\wedge d\theta^{\alpha 1} $.}, and
$L^a$ is the (pull-back to the worldsheet of the) VA 1--form (\ref{La=}) of flat ${\cal N}=1$ $D=10$ superspace $\Sigma^{(10|16)}_L$. It is invariant under
($D=10$, ${\cal N}=1$)
 supersymmetry with constant fermionic spinor parameter $\epsilon^\alpha_L$,
\begin{eqnarray}
\label{susyXL}
\delta_{\epsilon}X_L^a=  i \delta_{\epsilon} \theta^{1}\sigma^a\theta^{1}\; , \qquad \delta_{\epsilon} \theta^{\alpha 1}= \epsilon^{ \alpha 1}
\; .
\end{eqnarray}

The PST Lagrangian form (\ref{L=PST}) includes a 1--form
\begin{eqnarray}
\label{v=da}
v=d\xi^m v_m \: , \qquad v_m ={\partial_ma(x) \over \sqrt{\partial a g\partial a } }\; , \qquad
\end{eqnarray}
constructed from an auxiliary scalar field $a=a(\xi)$ called the {\it PST scalar} \cite{Pasti:1995tn}. This  obeys a topological restriction $\partial a g\partial a :=\partial_m a g^{mn}(\xi) \partial_n a\not=0$ (better
$\partial_m a g^{mn}(\xi) \partial_n a> 0$ \cite{Bandos:2014bva}), where $g^{mn}=g^{mn}(\xi)$ is the inverse of the worldsheet metric $g_{mn}$. That is also used to define the contraction, e.g.
\begin{eqnarray}
\label{ivLa=}
i_vL^a = v^m L_m^a = v_ng^{nm}L_m^a\, ,
\quad i_vd\theta^{\alpha 1}= v_ng^{nm} \partial_m \theta^{\alpha 1}.
\end{eqnarray}

We prefer to construct the worldsheet metric $g_{mn}$
from worldsheet zweibein 1-forms
\begin{eqnarray}
&& g_{mn}(\xi )= {1\over 2}\left( e_{m}^{+} e_{n}^{-} +  e_{m}^{-} e_{n}^{+} \right)\; \quad  \\
\label{e++=def}
&& e^{+}=d\xi^m e_m^{+}=e^0+e^1 , \quad e^{-}=d\xi^m e_m^{-} = e^0-e^1,  \qquad
\end{eqnarray}
which are among the independent variables of our dynamical system.
Then the Hodge duality  $*$ can be defined as a simple operation on the zweibeine
 \begin{eqnarray}\label{gmn=ee}
*e^{+}=e^{+}\; , \qquad *e^{-}=-e^{-}\; ,\quad
\end{eqnarray}
so that, for $L^a=
  e^{+}L_{+}^a +  e^{-}L_{-}^a$, we have $*L^a =
  e^{+}L_{+}^a -  e^{-}L_{-}^a\;$ and
  \begin{eqnarray} \label{*La=}
 L^a- *L^a =2 e^{-}L_{-}^a\;  . \end{eqnarray}

Decomposing the 1-form $v$ (\ref{v=da}) on the zweibein basis,
$v=d\xi^mv_m = e^{+}v_{+}+ e^{-} v_{-}$, one
notices that
\begin{eqnarray}\label{4v+v-=1}
4v_{+} v_{-}=1 \; .\quad
\end{eqnarray}
This equation implies that neither of the components $v_{+}$ and $v_{-}$ can vanish thus reflecting  the topological restriction on the PST scalar ($v_{+}  = {1\over 2}\sqrt{{\nabla_{+} a\over {\nabla}_{-}a}}= {1\over 4v_{-}}$).

The 'right' supermultiplet action $S_R= \int {\cal L}^R$ has the  similar structure
${\cal L}_{2}^R =\tilde{{\cal L}}_{2}^{PST}+  \tilde{{\cal L}}_{2}^{WZ}$ with
\begin{eqnarray}
\label{tL=PST}
 \tilde{{\cal L}}_{2}^{PST}&=& - R^a \wedge v \, i_v(R_a +*R_a)\; , \qquad \\
\label{tL=WZ} \tilde{{\cal L}}_{2}^{WZ} &=& + i dX_R^a \wedge  d\theta^2\sigma_a\theta^2 \;  \qquad
\end{eqnarray}
and  $R^a$ given in  (\ref{La=}). Notice that  we use the same zweibein (encoded in $*$)
and PST scalar (encoded in $v$ (\ref{v=da})) in both left and right parts of the action.

\section{Supersymmetry and gauge symmetries of the action}

\subsection{Supersymmetry}

Taking into account the famous D=10 Fierz identities $\sigma_{a \alpha(\beta}\sigma^{a}{}_{\gamma\delta)}\equiv 0$, we can write the (formal) exterior derivative of the WZ term, $d {\cal L}_{2}^{WZ} = - i dX_L^a \wedge d\theta^1\sigma_a\wedge d\theta^1$, in the form
\begin{eqnarray}
\label{dL=WZ}
d {\cal L}_{2}^{WZ} = - i L_a \wedge d\theta^1\sigma^a\wedge d\theta^1
\;
\end{eqnarray}
which makes manifest its invariance under  (\ref{susyXL}).
As ${\cal L}^{PST}_2$ in (\ref{L=PST}) is also invariant,   the action for the left supermultiplet, $S_L=\int {\cal L}^L_2$ (\ref{SL=}), is supersymmetric up to  integral of total derivative, which vanishes in the case of closed string.
The complete action (\ref{S=SL+SR}) is invariant  under  (\ref{susyXL}) as well as under  its counterpart acting on ${\cal Z}_R^{{M}}(\xi)$,
\begin{eqnarray}
\label{susyXR}
\delta_{\epsilon}X_R^a= i \delta_{\epsilon} \theta^{2}\sigma^a\theta^{2}\; , \qquad \delta_{\epsilon} \theta^{\alpha 2}_L= \epsilon^{\alpha 2}
\; , \qquad
\end{eqnarray}
so that the total number of target (super)space supersymmetries of our action is 32, the same as for type II GS superstrings.

\subsection{$\kappa$--symmetry}

The action is also invariant under 16=8+8 parametric local fermionic $\kappa$--symmetry. It is infinitely reducible: we describe it in terms of  32=16+16 fermionic functions $\kappa^{+}_\beta (\xi)$ and $\kappa^{-}_\beta (\xi)$ only half of which does contribute efficiently in the transformations of physical fields (chiral supermultiplets).
One half of the $\kappa$ symmetry with parametric functions $\kappa^{+}_\beta (\xi)$ acts on left coordinate functions and $e^{+}$,
\begin{eqnarray}
\label{kappa=symmL}
 \delta_{\kappa} \theta^{\alpha 1}= (L_{+}^a - 4(v_{+})^2 L_{-}^a  ) \tilde{\sigma}{}^{\alpha\beta}_a \kappa^{+}_\beta\; ,   \qquad \nonumber \\ \delta_{\kappa} X_L^a = i \delta_{\kappa} \theta^{1}\sigma^a\theta^{1}\; ,  \qquad   \qquad \nonumber \\ \label{kappa=symme+}  \delta_{\kappa} e^{+}= -4i e^{-}\nabla_{-}\theta^{\alpha 1}  \kappa_{\alpha}^{+} \; ,
\end{eqnarray}
while
the other  half  with parametric functions $\kappa^{-}_\beta (\xi)$ acts on left coordinate functions and $e^{-}$,
\begin{eqnarray} \label{kappa=symme-}
\label{kappa=symmR} \delta_{\kappa} \theta^{\alpha 2}= (R_{-}^a - 4(v_{-})^2 R_{+}^a  ) \tilde{\sigma}{}^{\alpha\beta}_a \kappa^{-}_\beta\; ,   \qquad \nonumber \\
\delta_{\kappa} X_R^a = i\delta_{\kappa} \theta^{2}\sigma^a\theta^{2}  \; ,  \qquad  \qquad   \nonumber \\ \delta_{\kappa} e^{-}= -4ie^{+} \nabla_{+}\theta^{\alpha 2}  \kappa_{\alpha}^{-} \; .
\end{eqnarray}
The PST scalar is inert under the $\kappa$--symmetry.

Notice that the $\kappa$--symmetry transformations for physical fields can be also written without explicit use of zweibeine,
\begin{eqnarray} \label{kappa=*}
\label{kappa=symmR}
\delta_{\kappa} \theta^{\alpha 1}= i_v*L^a \tilde{\sigma}{}^{\alpha\beta}_a \kappa^{1}_\beta\, , \qquad \delta_{\kappa} \theta^{\tilde{\alpha} 2}= i_v*R^a  \tilde{\sigma}{}^{\tilde{\alpha}\tilde{\beta}}_a \kappa^{2}_{\tilde{\beta}}\, ,  \quad
 \end{eqnarray}  with   $\kappa^{1}_\beta \propto  v_{+}  \kappa^{+}_\beta$ and $\kappa^{2}_\beta \propto  v_{-}  \kappa^{-}_\beta$.

The presence of $\kappa$--symmetry (first discovered in \cite{de Azcarraga:1982dw,Siegel:1983hh} for superparticle models) is important because it implies that the ground state of our dynamical system preserves some amount (1/2) of the supersymmetry (and is a stable, BPS state)
\cite{Bergshoeff:1997kr,Bandos:2001jx}. In our case it is also important to establish the relation of our model with type II GS superstring, which is known to possess the $\kappa$--symmetry reducing by half the number of its fermionic degrees of freedom.

\subsection{PST gauge symmetries }

Our action is invariant under gauge symmetry `parametrized' by an arbitrary (up to topological restrictions) variation of the PST scalar, $\delta a(\xi)$, supplemented by the following variation of the bosonic coordinate functions
\begin{eqnarray}
\label{PST2=symm}
 \delta_{_{PST_2}} X_L^b = {\cal L}^b \, \delta a(\xi)  \; , \qquad \delta_{_{PST_2}} X_R^b =  {\cal R}^b
\,   \delta a(\xi) , \quad
\\
\label{cLa=}
 {\cal L}^b:=  {i_v (L^b-*L_b)\over \sqrt{\partial a g\partial a}}\; , \qquad  {\cal R}^b := {i_v (R^b+*R_b)\over \sqrt{\partial a g\partial a}}\; .
\,    \quad
\end{eqnarray}
All other fields are inert under this   {\it PST$_2$ symmetry}, $\delta_{_{PST_2}} (^{other}_{fields})=0$, which  makes the PST scalar  a pure gauge,  St\"{u}ckelberg field, thus justifying our  statement on its auxiliary nature.   The topological restrictions $\partial a  g\partial a  \not= 0$ do not allow one to set $a(x)$ equal to constant or identify it with light--like coordinates, but one can fix e.g. the gauge $a(\xi)=\xi^0=\tau$. In this gauge our PST action acquires Floreanini--Jackiw-- or Henneaux--Teitelboim--like form (see \cite{Floreanini:1987as,Henneaux:1987hz}) in which its reparametrization (2d general coordinate) invariance is not manifest.

Anther important symmetry, which we call PST$_1$ symmetry (see \cite{Floreanini:1987as} for its non-covariant form), acts nontrivially only on the  bosonic coordinate functions,
\begin{eqnarray}
\label{PST1=symm}
 \delta_{_{PST_1}} X_L^a = \varphi^a (a(\xi)) \; , \qquad  \delta_{_{PST_1}}  X_R^a =  \tilde{\varphi}^a (a(\xi)) , \\ \nonumber    \delta_{_{PST_1}}  (^{other}_{fields})=0 \; ,  \qquad
\end{eqnarray}
and is characterized by 20=10+10 arbitrary functions of the PST scalar,  ${\varphi}^a (a(\xi))$ and $ \tilde{\varphi}^a (a(\xi))$. When
 $\partial a  g\partial a  > 0$ this infinite dimensional  symmetry can be shown to be the gauge symmetry so that it can be used to reduce the number of  degrees of freedom of  the dynamical system (see
 \cite{Bandos:2014bva} and refs. therein).

\section{Equations of motion}

The variation of the action (\ref{S=SL+SR})- (\ref{L=WZ}), (\ref{tL=PST}), (\ref{tL=WZ}) with respect to zweibein gives a beautiful deformed version of the Virasoro conditions
\begin{eqnarray}
\label{Eqe=Vir}
\left(L^a_+ -4(v_+)^2 L^a_-\right)^2=0\; , \qquad \left(R^a_- -4(v_-)^2 R^a_+\right)^2=0\; . \qquad
\end{eqnarray}
With this in mind, the form of the fermionic equations of motion,
\begin{eqnarray}
\label{Eq=--th1}
\nabla_- \theta^{\alpha 1}\sigma^a_{\alpha\beta} \, \left(L^a_+ -4(v_+)^2 L^a_-\right) =0  \, , \qquad \\ \label{Eq=++th2}
\nabla_{+} \theta^{\alpha 2}\sigma^a_{\alpha\beta} \, \left(R^a_- -4(v_-)^2 R^a_{+}\right) =0  \, , \qquad
\end{eqnarray}
(where $\nabla_- = e_-^m\partial_m$,  $  \nabla_{+}=e_{+}^m\partial_m$)
suggests the presence of the local fermionic $\kappa$--symmetry (\ref{kappa=symme+}), (\ref{kappa=symme-}) \footnote{Indeed, on the surface of the deformed Virasoro conditions (\ref{Eqe=Vir}), the contractions of  (\ref{Eq=--th1}) with $\tilde{\sigma}{}^{\beta\gamma}_b \, \left(L^b_+ -4(v_+)^2 L^b_-\right) $
and of (\ref{Eq=++th2}) with $\tilde{\sigma}{}^{\beta\gamma}_b \, \left(R^b_- -4(v_-)^2 R^b_+\right) $  vanish identically. These are the Noether identities for the $\kappa$--symmetry.}.

The variation with respect to the bosonic coordinate functions result in the second order equations
\begin{eqnarray}
\label{EqXL=}
d(da {\cal L}_a)=0\; , \qquad d(da {\cal R}_a)=0\; , \qquad
\end{eqnarray}
which imply that the ${\cal L}^a$ and ${\cal R}^a$ from (\ref{cLa=}) are equal to arbitrary functions of the PST scalar,  ${\cal L}^a= {\phi}^a (a(\xi))$ and ${\cal R}^a=\tilde{\phi}^a (a(\xi))$.
However, the arbitrary r.h.s. of these equations can be gauged away using the  PST$_1$ gauge symmetry (with
${d \varphi^b(a)\over da}=- \phi^b(a)$ and ${d\tilde{\varphi}{}^b(a)\over da}= - \tilde{\phi}^b(a)$). As a result, the gauge fixed form of the  Lagrangian  equations (\ref{EqXL=}) is the pair of homogeneous first order equations ${\cal L}^a= 0$ and ${\cal R}^a=0$. These, in their turn,  imply (see \cite{Pasti:1995tn} and e.g.   \cite{Bandos:2014bva}) the self-duality and anti-self-duality conditions
\begin{eqnarray}
\label{Eq=self-d}
L^a-*L^a=0 \, , \qquad R^a+*R^a=0 \; .  \qquad
\end{eqnarray}
In zweibein formalism these can be written in the form $L^a=e^+ L_+$ and $R^a=e^-R_-$ or
\begin{eqnarray}
\label{L--a=0}
L_-^a : =\nabla_- X_L^a - i \nabla_-\theta^1\sigma^a\theta^1 =0 \, , \qquad \\ \label{R++a=0} R_{+}^a:=\nabla_{+} X_R^a - i \nabla_{+} \theta^2\sigma^a\theta^2 =0  \; . \qquad
\end{eqnarray}
These equations provide the supersymmetric generalization of the chirality conditions characteristic for the type II GS superstring, in which case however, the first terms would involve the same bosonic coordinate field $X^a$.  Eqs. (\ref{L--a=0}) and (\ref{R++a=0}) also
make manifest that the deformed Virasoro conditions (\ref{Eqe=Vir}) reduce to the standard  $(L^a_+ )^2=0$ and $(R^a_- )^2=0$ in the 'self-duality gauge' (\ref{Eq=self-d}). The fermionic equations (\ref{Eq=--th1}) and (\ref{Eq=++th2}) in this gauge also reduce to  more standard  $\nabla_- \theta^{\alpha 1}\sigma^a_{\alpha\beta}  L^a_+ =0 $ $\nabla_{+} \theta^{\alpha 2}\sigma^a_{\alpha\beta} R^a_- =0 $.

Fixing the conformal gauge $e^+ =\propto d\xi^+$, $e^- = \propto d\xi^-$ (in which
$\nabla_-=\propto \partial_-$ and $  \nabla_{+}=\propto  \partial_{+}$), the  light cone gauge $X_L^+=\xi^+$, $X^-_R= \xi^-$, and the $\kappa$--symmetry gauge
\begin{eqnarray}
\label{kappa=gauge}
\sigma^+ \theta^1=0 \quad & \Leftrightarrow & \quad  \theta^{\alpha 1}= (0,\theta^{1-}_{\dot{q}}), \quad \dot{q}=1,...,8, \quad \nonumber \\  \sigma^- \theta^2=0  \quad & \Leftrightarrow & \quad    \theta^{\alpha 2}= (\theta^{2+}_{{q}}, 0) , \quad  {q}=1,...,8, \quad
\end{eqnarray}
we find that the equations of motion for physical fields are just the chirality and anti-chirality conditions
 \begin{eqnarray}
\label{d--XL=0}
 && \partial_-  X_L^I =0  , \quad \partial_- \theta^{1-}_{\dot{q}} =0 , \quad I=1,...,8\; ,  \quad \dot{q}=1,...,8,   \nonumber \\
 && \partial_{+} X_R^I=0,  \quad \partial_{+} \theta^{2+}_{{q}} =0  , \qquad \dot{q}=1,...,8\; . \qquad
\end{eqnarray}
Thus $( X_L^I ,  \theta^{1-}_{\dot{q}} )$ and $( X_L^I, \theta^{2+}_{{q}})$ can be identified with the physical left-moving and right-moving fields of the type II GS superstring. As $\partial_+X_L^-$ and $\partial_-X_R^+$ are  expressed through the above physical fields by
the solution of the Virasoro conditions, we
can  state that our superstring in double ${\cal N}=1$ superspace provides a  T-duality invariant formulation of the GS superstring. The details on T-duality symmetry of our action and on zero modes of chiral bosons will be discussed elsewhere.

\section{Conclusion and Outlook }

In this letter we have presented an action for the string in doubled ${\cal N}=1$ $D=10$ superspace
which provides a covariant and supersymmetric generalization of the Tseytlin's duality-symmetric action for bosonic string \cite{Tseytlin:1990nb}. We have shown that our action is invariant under a local fermionic $\kappa$--symmetry similar (but not identical) to the one of the standard GS superstring action. Our model uses essentially the PST approach to Lagrangian description of self-dual gauge fields, the set of which include chiral scalars (left-moving or right-moving fields) as $d=2$ representative. The `generalization' of our model for the case of superstring in doubled ${\cal N}=1$ superspace with $D=3,4$ and $6$ is straightforward.

The bosonic limit of our construction  gives the PST-type action for  bosonic string in doubled spacetime  constructed in \cite{DeAngelis:2013wba} so that the nontrivial part of our study was to prove that our supersymmetric generalization  does possess a local fermionic $\kappa$--symmetry which guarantees that the ground state of our dynamical system is 1/2 BPS state (see \cite{Bergshoeff:1997kr,Bandos:2001jx}) and allow to consider  our superstring in doubled superspace  as T-duality invariant formulation of the type II Green-Schwarz (GS) superstring.

It will be interesting to study the relation of our model to  the T-duality invariant superstring formulation recently proposed in \cite{Hatsuda:2015cia}. We see an advantage of our construction  in that it seems to be much more  economic: in contrast to  \cite{Hatsuda:2015cia}, we do not need either additional coordinates  (tensorial etc.) or an embedding  of a direct product of two super-Poincar\'e algebras into a superalgebra with non--degenerate metric. Probably the use of PST technique allowed for such a simplifications.

A natural application of our study is to consider the generalization of our action to curved doubled superspace and to use requirement of the preservation of the $\kappa$--symmetry to search for  constraints
defining supergravity in doubled superspace. Probably such a search for doubled supergravity  would shed  a light on the nature of section conditions which have to be imposed  by hand in doubled field theory as it is formulated presently; an even  more optimistic hope is that such an approach might generate an alternative
to the section conditions.

\acknowledgements{
This work has been supported in part by the Spanish MINECO grant FPA2012-35043-C02-01  partially financed  with FEDER funds,  by the Basque Government research group grant ITT559-10 and the  Basque Country University program UFI 11/55. The author is thankful to  the Theoretical Department of CERN for hospitality and support of his  visit in spring of 2015,
and to Dima Sorokin for reading the manuscript and useful comments. }

\end{document}